\documentclass[aip,graphicx]{revtex4-1}
\usepackage{graphicx}
\usepackage{dcolumn}
\usepackage{bm}
\usepackage{color}
\usepackage{soul} 

\usepackage[utf8]{inputenc}
\usepackage[T1]{fontenc}
\usepackage{mathptmx}
\usepackage{etoolbox}

\draft 

\begin{document}

\title{Exploring the Complex Landscape of Entropy Stabilized Oxides} 

\author{Bo Jiang*}
\email[]{boji@uio.no}
\affiliation{Department of Materials Science and Engineering, University of Tennessee, Knoxville, TN 37996, USA}
\affiliation{Neutron Scattering Division, Oak Ridge National Laboratory, Oak Ridge, TN 37830, USA}
\affiliation{Department of Chemistry, Centre for Materials Science and Nanotechnology, University of Oslo, POB 1033 Blindern, NO-0315 Oslo, Norway}

\author{De-Ye Lin}
\affiliation{CAEP software Center for High Performance Numercial Simulation, Huayuan road 6, Beijing 100088, China}
\affiliation{Institute of Applied Physics and Computational Mathematics, Huayuan road 6, Beijing 100088, China}

\author{Gerald R. Bejger}
\affiliation{Department of Physics and Astronomy, James Madison University, Harrisonburg, Virginia 22807, USA}
\affiliation{Department of Materials Science and Engineering, Virginia Polytechnic and State University, Blacksburg, VA 24061, USA}

\author{Stephen C. Purdy}
\affiliation{Manufacturing Science Division, Oak Ridge National Laboratory, Oak Ridge, TN 37830, USA}

\author{Yuanpeng Zhang}
\affiliation{Neutron Scattering Division, Oak Ridge National Laboratory, Oak Ridge, TN 37830, USA}

\author{Xin Wang}
\affiliation{Department of Materials Science and Engineering, University of Tennessee, Knoxville, TN 37996, USA}

\author{Jon-Paul Maria}
\affiliation{Department of Materials Science and Engineering, The Pennsylvania State University, University Park, PA 16802, USA}

\author{Christina M. Rost*}
\email[]{cmrost@vt.edu}
\affiliation{Department of Physics and Astronomy, James Madison University, Harrisonburg, Virginia 22807, USA}
\affiliation{Department of Materials Science and Engineering, Virginia Polytechnic and State University, Blacksburg, VA 24061, USA}

\author{Katharine Page*}
\email[]{kpage10@utk.edu}
\affiliation{Department of Materials Science and Engineering, University of Tennessee, Knoxville, TN 37996, USA}
\affiliation{Neutron Scattering Division, Oak Ridge National Laboratory, Oak Ridge, TN 37830, USA}


\begin{abstract}
 Entropy-stabilized oxides (ESOs), driven by high configurational entropy, have gained phenomenological research interest due to their potential for tailoring  structure-property relationships. However, the chemical short-range ordering (SRO) and its interplay with local lattice distortion (LD) remain to be explored, although they could diminish the configurational entropy and potentially impact structure-property relationships. A combination of experimental  and theoretical approaches are employed to investigate the SRO and LD in the prototype ESO, (Mg$_{0.2}$Co$_{0.2}$Ni$_{0.2}$Cu$_{0.2}$Zn$_{0.2}$)O, generally referred to as J14. We demonstrate that the efficiency and accuracy of density functional theory (DFT) relaxed special quasirandom structures (SQS) enhances the analysis of the local structure of J14, unveiling the unique local cationic environments. Importantly, this joint experimental and computational approach sheds light on the understanding of local structure and structure–property relationships in J14, demonstrating the necessity for further research into other high entropy and compositionally complex materials.

\end{abstract}


\maketitle 

\section{\label{sec:level1}Introduction}
Entropy-stabilized oxides (ESO), first reported within J14 (Mg$_{0.2}$Co$_{0.2}$Ni$_{0.2}$Cu$_{0.2}$Zn$_{0.2}$)O by Rost et. al in 2015, represent the most strictly defined subset of high entropy oxide (HEO) and compositionally complex oxides (CCO) based on extension of concepts first reported in high-entropy alloys (HEAs).\cite{rost2015entropy,oses2020high, brahlek2022name}
J14 composition belongs to a class of rock salt compounds created by purposefully adding five or more cations to a single crystallographic lattice site, which exhibit unusual thermodynamic and electronic properties due to the presence of large configurational entropy with a high degree of disorder.\cite{musico2020emergent,oses2020high, sarkar2020high,fracchia2022configurational,spurling2022entropy,pu2023mg}
The entropy associated with this disorder can stabilize the material against phase transitions or other changes at high temperatures, leading to unique properties such as high thermal conductivity, superconductivity, and ionic conductivity in electrochemical energy storage applications.\cite{sarkar2018high,zhao2020high,sun2021high,ma2021high,salian2022entropy,zeng2022high,aamlid2023understanding} However, the impact of local structure or the degree of configurational disorder/entropy on the stabilization of ESO remains unclear, presenting a challenge in the exploration of structure-property relationships.

Chemical short-range ordering (SRO) and local lattice distortion (LD) are believed to be major contributions to macroscopic properties in solid solutions with  multiple elements.\cite{dragoe2019order,zhang2017local,fantin2020short} Because of extremely challenging  deciphering of SRO for HEAs so far,\cite{chen2021direct} let alone J14, a full scope of the structure-property trends of high entropy systems have yet to be understood. 
Determining the local structure of compositionally disordered systems such as high entropy materials requires theoretical modeling methods and the use of multiple characterization techniques to gain a reasonable understanding of ordering at different length scales.
Density functional theory (DFT) has been widely used in the literature to aid in both property prediction and structural analysis of HEAs. DFT has been used to generate structure models and to help explain the magnetic properties of high entropy materials. \cite{rak2018evidence,rak2020exchange,jiang2020probing} 
Monte Carlo (MC) methods have also been used, namely reverse Monte Carlo (RMC) and metropolis Monte Carlo (MMC), to determine local ordering and phase fraction, respectively, in neutron total scattering experiments.\cite{pitike2020predicting, jiang2020probing,jiang2023local} 
Classic molecular dynamics (MD) simulations have been performed for calculating phonon-mediated thermal conductivity and atom type distorts, and further ab initio molecular dynamics (AIMD) simulations confirm Jahn-Teller distortions around specific cation species.\cite{lim2019influence,kaufman2021tunable,jiang2023local} 

Several experimental attempts have been pursued to scrutinize the local structure of J14. 
The use of X-ray diffraction (XRD) and Rietveld refinement has been employed previously to determine the lattice parameter changes in J14 and other high entropy materials.\cite{zhang2020applying,usharani2020antiferromagnetism}  XRD provides information about long-range crystallographic ordering based on families of planes, and when working with elements of similar scattering cross-sections, can lead to uncertainties in the short range. Additional characterization techniques, such as scanning electron microscopy (SEM), energy dispersive spectroscopy (EDS), and atom probe tomography (APT), have also been employed in J14. \cite{rost2015entropy,rost2017local,chellali2019homogeneity,dupuy2021multiscale,jothi2022persistent}  Once an atomic distribution has been confirmed, this enables a basic starting point in which local structure techniques can be employed and analyzed.

However, most of these experimental studies are limited to a cursory understanding of local structural information due to technical limitations. It is worthwhile to state that X-ray absorption spectroscopy (XAS),\cite{rost2017local, harrington2019phase,johnstone2022entropy,braun2018charge,pu2023mg} neutron diffraction, and  pair distribution  function analysis (total scattering)  have been used to elucidate local structure details of complex materials.\cite{zhang2019long,jiang2020probing} 
Rost et al. demonstrated the use of extended X-ray absorption fine structure (EXAFS) to determine the average local environment of the component cations in J14 and found Cu to exhibit a well-known Jahn-Teller distortion.\cite{rost2017local} However, despite the collection of both synchrotron and neutron total scattering data, the challenges in fitting the data effectively still have to be resolved.

Recently, total scattering studies and especially the neutron have been used to probe the local structure and dynamics of HEOs,\cite{jiang2020probing} providing insight into the nature of the disorder present in HEO materials and helping to understand the underlying mechanisms that give rise to their unique properties. Few classical MD simulation studies have been reported on ESOs and even HEOs, due to the absence of accurate potentials that can describe the complex interactions within them.\cite{akrami2021high} Special quasirandom structures (SQS) combined with DFT and AIMD have demonstrated the capability to predict local and global properties of disordered HEAs,\cite{santodonato2015deviation} as SQSs have been shown to closely mimic the true randomness in complex structures.  Prior research has demonstrated the effective use of SQS-DFT method in conjunction with total scattering data to analyze local structure in ferroelectric solid solutions and pyrochlore HEOs.\cite{jiang2020probing,jiang2022special,tomboc2022stabilization}
With the combination of these techniques, there are still unresolved characterization challenges associated with modeling techniques and modeling parameters, such as distinguishing structural traits specific to HEOs.\cite{jiang2023local}

Structural models of highly disordered systems that encompass length scales across SRO boundaries are difficult to develop due to the inherent complexity that comes with multiple cations occupying the same crystallographic site. This is an issue when it comes to local structure analysis as characterization techniques, such as EXAFS, generally use a unit cell- based (`small-box') structure model to fit and interpret the data. Incorporating techniques such as large-box RMC to fit neutron total scattering and EXAFS spectra simultaneously enables a more accurate model representation. A complete analysis of J14 using multiple theoretical and experimental methods to confirm the local structure and local distortions that exist within J14 has yet to be done simultaneously.
In this study, we conducted an experimental and theoretical examination of the degree of SRO and LD in J14. The impact of SQS models with different sizes on the final results has been evaluated through this work. Additionally, the `small-box' modeling approach was performed to study the LD with the assistance of the DFT relaxed SQS model. Furthermore, atomic distribution analysis through AIMD simulations was used to investigate the local chemical environment and interatomic interaction. More advanced analysis using `large-box' RMC simulations by fitting neutron total scattering and EXAFS data simultaneously are performed to explore possible local ordering of cations in J14. 



\section{Results and Discussion}
\begin{figure*}
\includegraphics[width=\linewidth]{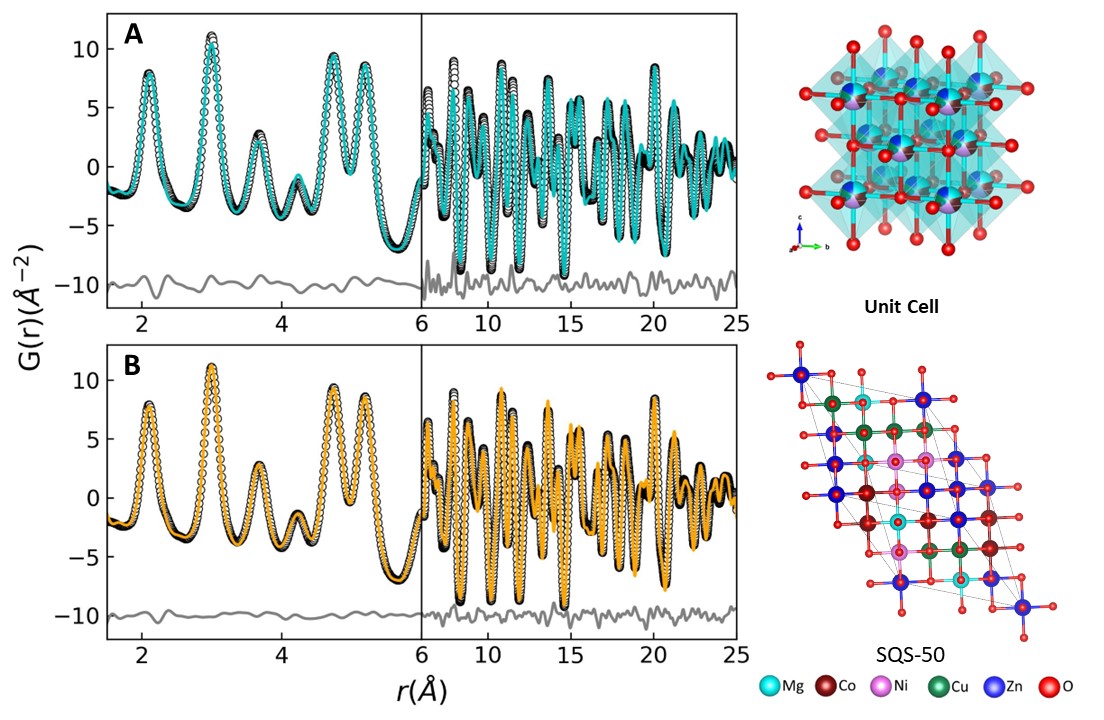}
\caption{\label{fig1:model} \textbf{Neutron total scattering data analysis.} Results of `small-box' refinements of neutron PDFs over local and medium range to (\textbf{A}) disordered mixed A-site rock-salt unitcell and (\textbf{B}) 50  atoms SQS for J14, with the corresponding structure models shown on right.}
\label{fig1}
\end{figure*}

The neutron diffraction Rietveld refinement and total scattering PDF analysis for the J14 sample with space group $Fm\overline{3}m$ with disordered mixed A-site rocksalt has been described in previous work\cite{rost2015entropy,chellali2019homogeneity}. The results in Figure S2 show a fair goodness-of-fit $R_{wp}$ = 7.44\% to experimental neutron diffraction data, but a poor fit to neutron PDF in Figure\,\ref{fig1}A, especially in the local $r$ range for the first peak. This indicates that the local structure cannot be simply described by a disordered cubic structure with mixed A-site. Driven by our prior experience with other high-entropy oxides\cite{jiang2020probing,jiang2023local}, the SQS model with different cell sizes has been designed to evaluate the feasibility of mimicking random structure in J14, as shown in Figure S3. It is crucial to emphasize at this point that improving computational effectiveness and SQS cell size is necessary to achieve accurate and reliable results.

One criterion used to evaluate the cation ordering in the SQS model of ESOs is the nearest neighbor function $n_{ij}(r)$ \cite{keen2001comparison,hui2007neutron}, which is defined as the mean number of atoms $i$ surrounding a central atom $j$;  \( n_{ij} \left( r \right) = \int _{r2}^{r1}4 \pi r^{2}c_{j} \rho _{0}g_{ij} \left( r \right) dr \).  Here $c_j$ is the proportion of atom $j$ in the compound, $\rho_0$ is the average number density of the compound, and $g_{ij}$ is the partial PDF for atoms $i$ and $j$. Figure S3 shows the $n_{A1-A2}(r)$ of A$_1$-A$_2$ cation pairs derived from the SQS model with different cell sizes up to 80 atoms. It is clear to see that increased cell size for SQS would yield configurations that are closer to the real random disordered ESOs. Especially in the SQS model with 50 atoms, the first shell of nearest neighbors for A$_1$-A$_2$ cation pairs has the same values. As shown in Figure\,\ref{fig1}B, the neutron PDF data can be well fitted by DFT relaxed 50 atoms SQS. The cation positions of the generated 50 cations SQS are given in Table
S1. However, due to the limited size of the supercell by the computational expenses of DFT, a perfect SQS for J14 may never be found. 
The larger SQS cell is better for reproducing the correlation functions of a random cation distribution in J14, but it is still worth exploring the limits of smaller SQS. As seen in Figure S4, we compare the PDF fits from large supercells with 250 atoms using both SQS and random distribution methods. As expected, the large SQS model with 250 atoms fits the neutron PDF very well from local 1.5-10 {\AA} to medium range up to 30 {\AA}. Although the local structure  1.5-10 {\AA} fits also well to a large random distribution model, the whole range 1.5-30 {\AA} fits show obvious differences.  This also proves near complete chemical homogeneity distribution of elements in J14 from the other side. These SQS models with 10 to 80 atoms were fully relaxed by DFT, and subsequently used to fit the PDF data in PDFgui with results shown in Figure S5. The successful decomposition of the partial PDFs from individual A-O pair-pair correlations is shown below for comparison. It should be noted that besides the SQS with 10 atoms model, the Cu$^{2+}$ from other models exhibited JT distortion in an octahedral environment with asymmetry Cu-O bond peaks at around 2.1 {\AA} consisting of four planar bonds and two axial bonds.\cite{rost2017local}

\begin{figure}[htbp]
\includegraphics[width=\linewidth]{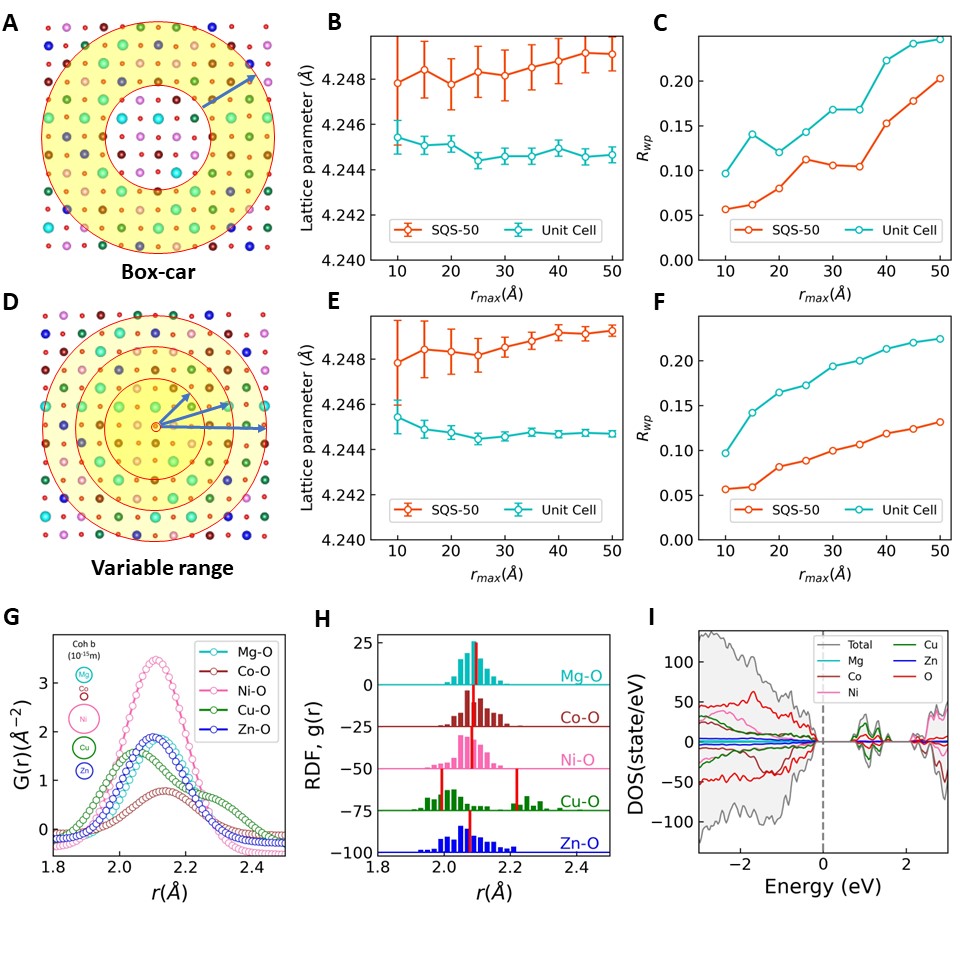}
\caption{\label{fig2:PDFgui}\textbf{PDF refinements and DFT calculations.} The (\textbf{A}) box-car and (\textbf{D}) varying $r_{max}$ range `small-box' refinements results of neutron PDFs of (\textbf{B, E}) lattice parameters and (\textbf{C, F})  $R_{wp}$ values using both disordered mixed A-site unit cell and SQS with 50 atoms models. (\textbf{G}) Decomposed partial PDFs for the first A-O peaks computed from `small-box' refinements using SQS with 50 atoms model. Insert shows the coherent neutron scattering  lengths (scattering factors) for different elements are drawn as circles to scale. (\textbf{H}) shows the histograms of sum bond lengths on the M-cations sites from five DFT relaxed SQSs with 50 atoms with different A-sites sorting order (s1 to s5 as shown in Figure S6). The experimentally bond length from EXAFS is indicated by the red solid line. (\textbf{I}) Total and partial electronic density of states of J14 from DFT simulations with 250 atoms supercell from SQS.  }
\label{fig2}
\end{figure}

Based on our previous experience,\cite{jiang2021effect,jiang2022temperature} it is common to determine a length scale for local atomic departures from typical structure motifs using two distinct `small-box' PDFs sequentially refinements methods, the box-car, and varying $r$-range techniques.\cite{culbertson2020neutron,hou2018local} Box-car refinements were performed using $r$-ranges of 1-10, 5-15...40-50 {\AA} with identical length (Figure\,\ref{fig2}A), and varying $r$-range refinements were carried out with a fixed initial $r_{min}$ at 1 {\AA} with varying the $r_{max}$ from 10 to 50 {\AA} (Figure\,\ref{fig2}D). The evolution of lattice parameters and goodness of fitting $R_{wp}$ values with both methods are shown in Figure\,\ref{fig2}(B, C) and Figure\,\ref{fig2}(E, F), respectively. The results of both methods regarding lattice parameters are quite similar. While the lattice parameters produced from both methods are slightly larger than the long-range lattice parameter $\sim$ 4.24 {\AA} from Rietveld refinements. However, the SQS model shows a modest rise while the unit-cell model shows a minor decrease in lattice parameters. Note that the results from varying $r$-range refinements revealed oscillations in the low $r$ fitting range, which gradually stabilized as the fitting range increased. It is evident in Figure\,\ref{fig2}C and F that the SQS model significantly reduced the $R_{wp}$ value to almost half of the $R_{wp}$ value obtained from the unit cell model at low $r$ range, suggesting SQS model's capacity to describe structural features across a broad length scale from the local to the long-range scale.
This part concludes that these deviations between SQS and unit-cell models are directly correlated with the distinct local environments surrounding various cations, which indicate the considerable disorder from  local to long range and imply the presence of an estimated correlation length of LD of $\sim$2 nm.

In this manner, the partial PDFs that were successfully decomposed from the DFT relaxed SQS fits for individual cation-oxygen pairs were presented in Figure\,\ref{fig2}G, and the JT distortion of the Cu-O polyhedra can be observed. We also note the significant difference in the height of the first A-O peaks varies due to the neutron scattering length difference. To get more average results, we sort the cations by forcing each cation to move through the same places as every other cation, to produce five distinct configurations, and then fully relax using DFT, as denoted as s1 to s5 shown in Figure S6. We can see that although the initial local magnetic moments  were set as antiferromagnetic, the final results after DFT relaxation, showed local non-zero total magnetic moment and multiple random magnetic configurations. However, it is noteworthy that the arrangement of cations and magnetic moments will produce qualitatively similar outcomes,\cite{meisenheimer2019magnetic} as indicated by the minimal disparities in the fitting of local and medium-range using various sorting configurations, as demonstrated in Figure S7. The histograms of sum bond lengths from the above five configurations exhibit distinct bond distribution characteristics (Figure\,\ref{fig2}H), particularly the observable Cu-O bimodal distribution due to JT distortion. Note the peaks of Zn-O pairs exhibit a broader distribution compared to other pairs. However, all cation-oxygen pairs do not exhibit an ideal Gaussian distribution, which may be attributed to the limited number of available configurations. 
An analysis of the total and partial density of states (DOS) for DFT relaxed SQS models with 50 and 250 atoms is presented in Figures S8 and S9, respectively, for comparison. We observe an energy band gap (E$_g$, gap between the valence band maximum and conduction band minimum ) of $\sim$1.2 and $\sim$0.6 eV for two models, respectively. These values closely approximate the experimental band gap E$_g$ $\sim$0.8 eV. \cite{berardan2016colossal} Considering that DFT typically underestimates band gap, the larger E$_g$ obtained from small supercell can be related to the finite size effects. As evident from the partial DOS in Figure\,\ref{fig2}I (more details in Figure S9), it is noteworthy that the Cu-$d$ and Ni-$d$ orbital exhibit distinct distributions in both up and down spin due to the prevailing magnetism. Meanwhile, it is the Cu-$d$ and O-$p$ orbital that contribute most of the low-energy part of the valence band in both up and down spin. Overall, our analysis demonstrates the superiority of the SQS model in conjunction with DFT simulations to handle the compositional complexity in J14. Moreover, it is essential to conduct additional AIMD and complex RMC simulations to further examine and validate the model.


\begin{figure}[htbp]
\includegraphics[width=\linewidth]{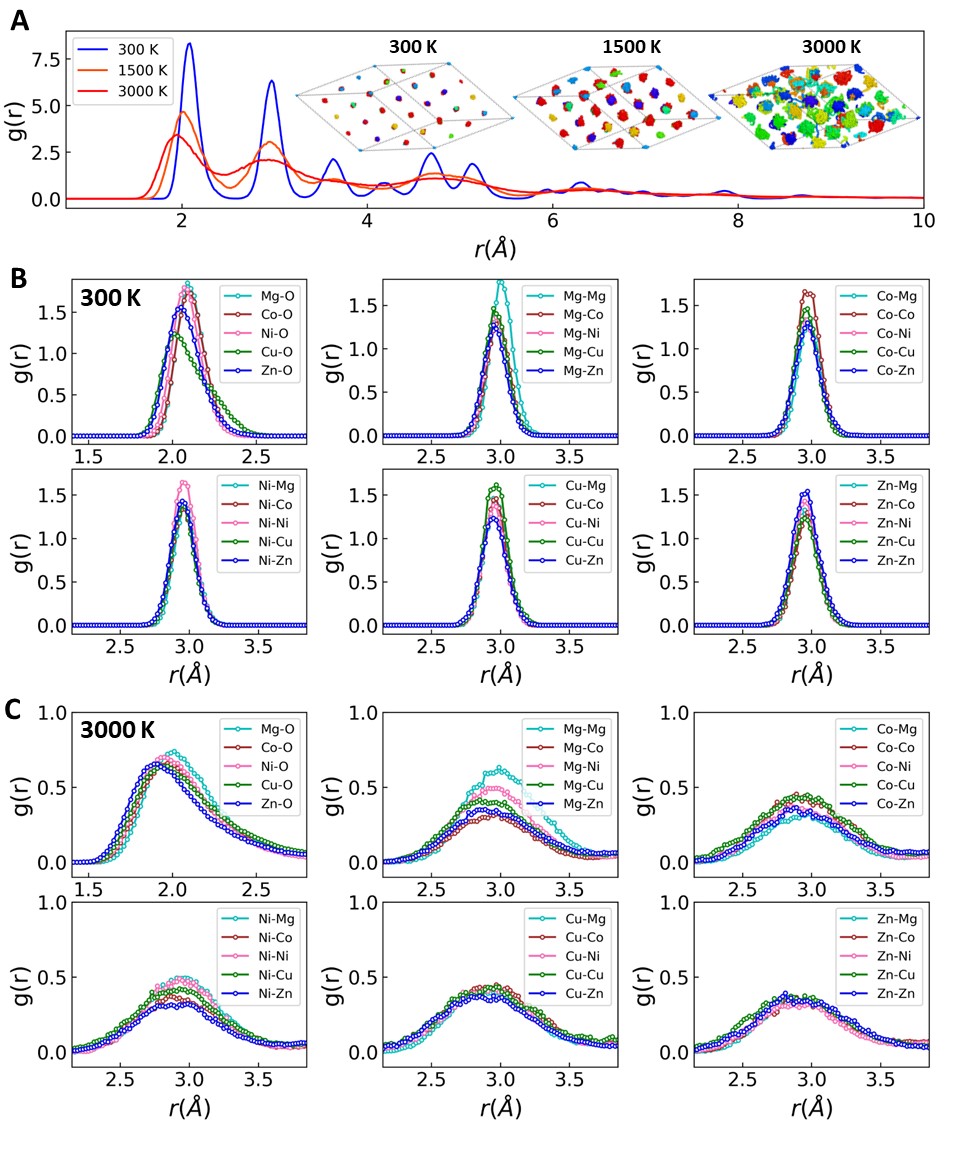}
\caption{\textbf{Tempearture dependence analysis from AIMD.} (\textbf{A}) Total radial distribution functions from AIMD simulations at 300 K, 1500 K and 3000 K using 50 atoms SQSs, inserts show the corresponding trajectory projections with color mapping. Plots of radial distribution functions of A-O and A-A bonds from AIMD simulation at (\textbf{B}) 300 K and (\textbf{C}) 3000 K.}
\label{fig3}
\end{figure}

Since the accurate interatomic potential and parameters using classical MD for ESO have not yet been established,\cite{anand2018phase,giri2017minimum, akrami2021high} the more precise but computationally costly AIMD simulations were carried out here to investigate the dynamical characteristics of J14. The total radial distribution functions (RDF) results from AIMD simulations from ambient temperature 300 K were shown in Figure\,\ref{fig3}A. The corresponding partial RDFs using 50 atoms SQSs with various configurations (s1 to s5 in Figure S6) are illustrated in Figure S10-S14, and the average partial RDFs results are illustrated in Figure\,\ref{fig3}B. Only the first nearest-neighbor peak is displayed for simplicity. Note that different configurations models from s1 to s5 exhibit distinct features of A-O and A-A bond distribution because of the arrangements of cations and magnetic moments. To avoid potential discrepancies in distinct SQS occupancy sites, we have ensured that each type of cation was thoroughly exchanged with the positions of other cations in configurations s1 to s5. Furthermore, the RDF from the AIMD simulation using the large SQS with 250 atoms has also been tested and did not exhibit any significant distinct features as seen in Figure S15. The average A-O pairs were found to be quite comparable to the results in Figure\,\ref{fig2}H. The Zn-O pair exhibited a lower height of peak and a broader distribution, which can be attributed to the shorter Zn-O bond length and the preference of ZnO for the wurtzite structure. Furthermore, there is a slightly preferred correlation for A-A pairs of identical cations, which is evident for all cations. These pair correlation features suggest the potential for trending to local clusters of identical atoms. Another noteworthy observation is the increased average of the Mg-Mg bond length, which can be attributed to the relatively larger radius of Mg atoms (1.6 {\AA}). To mitigate the influence of the low self-diffusion coefficient of atoms at ambient temperature, additional AIMD simulations were performed at elevated temperatures of 1500 K and 3000 K, as also depicted in Figure\,\ref{fig3}A, and the partial PDFs are shown in Figure S16 and Figure\,\ref{fig3}C. Note that both Structures at 300 K and 1500 K exhibit structured configurations, indicating that lattice features are preserved. However, the JT distortion of the Cu-O bond becomes barely noticeable at 1500 K. Aside from the first two peaks, the remaining peaks lose their lattice characters after 3000 K and shift into a condition of melting. Further analysis of the AIMD trajectory data at 3000 K also reveals that atoms deviate from their ordered lattice positions as shown in the insert in Figure\,\ref{fig3}A, displaying a loss of the solid structure. Examination of average partial RDFs results for 3000 K shows that all A-O pairs become blurred in Figure\,\ref{fig3}C. In contrast, the intensity of the Mg-Mg pair is significantly higher than that of other A-A pairs, even double that of the Mg-Co pair. This behavior can be explained by the smaller cations and the lightest atomic weight of Mg$^{2+}$, indicating a higher degree of diffusion compared to other cations at elevated temperatures.

\begin{figure}[htbp]
\includegraphics[width=\linewidth]{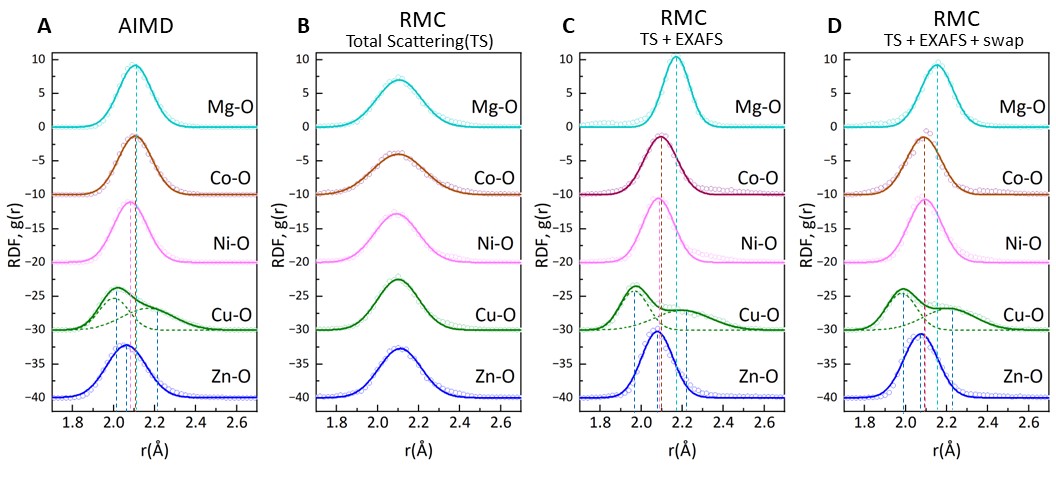}
\caption{\textbf{AIMD and RMC analysis.} Gaussian peak fitting of first A-O pair correlation from RDF of (\textbf{A}) AIMD simulation, (\textbf{B}) RMC result with fit to neutron total scattering data, (\textbf{C}) RMC result with fit to neutron total scattering and EXAFS data, (\textbf{D}) RMC result with fit to neutron total scattering and EXAFS data with atom-swapping, respectively. A doublet deconvoluted Gaussian peak was used  to fit Cu-O bond distribution peaks.}
\label{fig4}
\end{figure}

Simultaneous refinements of the local structure in J14 were conducted using neutron total scattering data, along with Co, Ni, Cu, and Zn K-edge EXAFS spectra, employing the latest RMCProfile v6 package.\cite{krayzman2009combined,zhang2020new} This approach has been found to be highly efficient for the investigation of disordered materials with mixed sites, as demonstrated in earlier studies.\cite{levin2019nanoscale,aoyagi2020controlling} In recent years, many attempts have been made to employ the RMC method for the analysis of SRO in HEAs from total scattering data.\cite{owen2016new,owen2017assessment,nygaard2021average} However, there is still limited literature that delves into the investigation of SRO in CCO/HEO/ESOs. In previous work on pyrochlore HEOs,\cite{jiang2020probing} 
we identified limitations of the RMC simulation in resolving a high number of similar cations systems with complicated M-O peak overlap and also proposed the possibility of introducing EXAFS spectra data into RMC fitting as a means of improving fitting. It is worth noting that the effectiveness of constraining RMC refinement with EXAFS relies on the k range, which means an $R$ space resolution of approximately 0.2 \AA in this study.
Figure S17 shows the RMC fitting results for all data sets including EXAFS FTs for Co, Ni, Cu, and Zn. We can see that the fitting to PDF exhibits high quality to the entire range. It should be noted that including the atom-swapping has limited improvements on the fitting (Figure S18), demonstrating the near complete chemical homogeneity at the local level for J14. The RMC fitting results are dependent on much of the amount of input data and potential constraints used. Nevertheless, the soft BVS constraint\cite{norberg2009bond} applied in RMC fitting is not effective in distinguishing the difference of atoms with similar properties in J14, such as those with similar atomic radii, a charge of +2, and even similar coherent neutron scattering lengths. Figure\,\ref{fig4} summarizes the first A-O pair correlations fitted to Gaussian curves for various models from AIMD and RMC fits. It is worth noting that the RMC fitting (without EXAFS data) produced the identical peak distribution for each A-O pair correlation, demonstrating the inability of RMC refinement total scattering alone to elucidate SRO and LD in J14.. The further incorporation of EXAFS data resulted in a notable improvement of the fitting results, which closely resembled the peak distribution obtained from AIMD. Due to the absence of EXAFS data for Mg, the peak of Mg-O pair correlation exhibited unreasonable high intensity (Figure\,\ref{fig4}C), which was further improved after introducing atoms-swapping constraint (Figure\,\ref{fig4}D). However, there is still a small deviation in the peak position of Mg-O pair correlations between the AIMD and RMC fits, which may result from the absence of EXAFS data and a larger radius of Mg atoms, further research on the behavior of Mg is beneficial and desirable.


\begin{figure}[htbp]
\includegraphics[width=\linewidth]{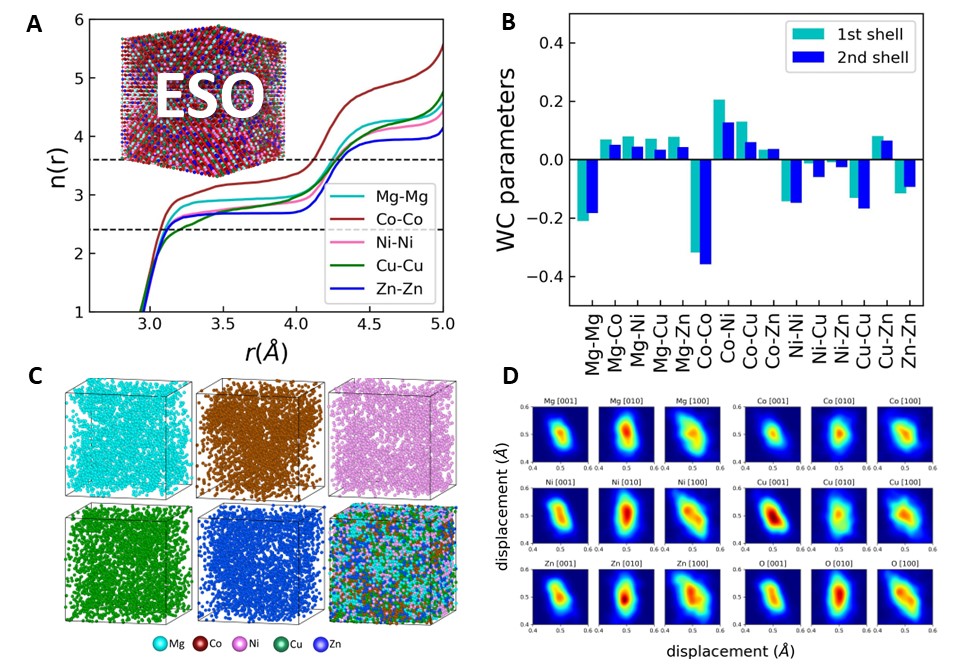}
\caption{\textbf{Local ordering and atom distribution.} (\textbf{A}) Nearest neighbor functions $n_{A-A}(r)$ indicating the local SRO of M-site cations from RMC configurations after atom-swapping. (\textbf{B}) The Warren-Cowley (WC) parameters of first shell and second neighbor correlations. (\textbf{C}) Cations and oxygen atoms distribution from refined RMC configurations. (\textbf{D}) Density plots with hot spots for all cations and oxygen from folded point clouds of RMC configuration are displayed along different directions.}
\label{fig5}
\end{figure}

We have recently discussed the challenge OF studying SRO in  previous pyrochlore and Ruddlesden–Popper (RP) CCO work.\cite{jiang2020probing,jiang2023local} Here, we examine all of the n$_{A-A}$(r) of A-A cation pairs derived from RMC simulations, with and without adding EXAFS spectra data, as shown in Figure S19 and S20, respectively. It is interesting to note that n$_{A-A}$(r) exhibits a larger value for identical cation pairs compared to those involving different cations, and similar results are observed after adding EXAFS data. The selected n$_{A-A}$(r) results for identical cation pairs and the corresponding Warren-Cowley (WC) parameters short-range order (SRO) parameters \cite{cowley1965short} for first- and second-neighbor correlations of all cation pairs are shown in Figure\,\ref{fig5}A and Figure\,\ref{fig5}B. We have extended these WC parameters to J14 as a useful tool for studying the SRO. However, the current definition of WC parameters for multi-component CCO/HEO/ESO is considered to be overly simplistic. A more accurate and appropriate definition is expected to gain insights understanding how different cations and other conditions affect the local ordering of cations. The n$_{A-A}$(r) and WC results are consistent with the RDF analysis of AIMD in Figure\,\ref{fig3}B. However, these results do not rule out the possibility of overfitting with a high number of cations involved. The cation distribution map of RMC supercell configuration was present in Figure\,\ref{fig5}C by colored dots. Even though, the intuitively compositional distribution of cations is visualized to be fairly homogeneous in macroscopic. We further folded the RMC supercell configuration into a single $Fm\overline{3}m$ unitcell to visualize the point `cloud' and compared it with the anisotropic atomic displacement parameters (ADP) obtained from `small-box' modeling, as shown in Figure S21. The oxygen `cloud' shows a broader distribution compared to the cation `cloud' sites. It is worth noting that the elongated ellipsoidal shape of the point `cloud' and ADP implies some degree of LD in the oxygen structure due to the different A-site cation environments. Further inspection of the hot density map of individual cation and oxygen distributions is shown in Figure\,\ref{fig5}D. It is interesting to notice that the Co cation clouds show more spheroidal distributions, and Ni/Zn cation clouds show a broader distribution due to smallest size of Ni and shortest bond of Zn (Mg$^{2+}$ 0.72 \AA, Co$^{2+}$ high-spin 0.745 \AA, Ni$^{2+}$ 0.69 \AA, Cu$^{2+}$ 0.73 \AA, and Zn$^{2+}$ 0.74 \AA)\cite{shannon1976revised} compared to other A site cations. 
We can conclude that the local structure of multi-component oxide material demonstrates tremendous complexity. Each cation strives to maintain the stability of its local environment, while also simultaneously coordinating with other cations to maintain the overall structural stability through entropy stabilization.


The coherence range of localized ordering of cations and defects can have profound impacts on properties in the simplest compositions.\cite{jorgensen1991defects} As compositional and consequently structural complexity is increased, the importance of systematic multi-length scale characterization and modeling must be stressed.  In this work, we combine the information gained from EXAFS, neutron scattering, and theoretical modeling to provide a complete picture of the disorder of HEO J14 across all length scales. As a next step, machine learning (ML)\cite{mobarak2023scope} enables the combination of research efforts in theory and advanced characterization to develop frameworks for complete structure analyses of high entropy materials. Already, ML has transformed materials science research from new materials design via the Materials Project\cite{jain2013materials}, property prediction\cite{ward2016general}, high throughput experimentation\cite{green2017fulfilling}, and Artificial Intelligence (AI)-assisted data interpretation as discussed in the review by Stein and Gregoire\cite{stein2019progress}. In the review article by Green et al.\cite{green2017fulfilling}, a key takeaway is the necessity for improved data libraries that include characterization beyond composition space. Currently, AI-assisted data analysis frameworks are restricted to individual characterization techniques, such as TEM\cite{wang2024artificial} or XRD\cite{maffettone2021crystallography}, which will likely have the same shortfalls of using isolated characterization methodologies discussed previously. We propose incorporating the combined results of this study into ML libraries as a starting point toward more rapid data interpretation of HEOs. Making such information and results openly available for AI learning will certainly increase the rate of progression of categorizing structure-property relationships in HEOs, a step toward realizing implementable applications.

\clearpage
\section{Conclusion}

In summary, we presented an experimental and theoretical investigation of the degree of SRO and LD in J14 ESO. Despite J14 being renowned for its multi-element single phase structure and increasing research interest, it remains challenging to detect the locally disordered cations within its solid solution. Through a comparison of the computed neutron PDFs and nearest-neighbor functions using supercells with various sizes, our simulations demonstrate that the SQS model with 50 atoms strikes an optimal balance between computational efficiency for DFT \& AIMD simulations and accurate results. 
Compared to a disordered mixed-cation unit cell model, the DFT relaxed SQS exhibits remarkable capability to improve the local and medium-range structure refinements. This effect holds even for large supercells with randomly distributed cations, indicating that the J14 could be regarded as possessing almost complete chemical homogeneity down to the atomic scale.
Individual partial PDFs were successfully decomposed from `small-box' modeling with the aid of DFT-relaxed SQS. Results from box-car refinements and varying $r$-range refinements indicate the ability of DFT relaxed SQS model to describe structural features across different length scales while accommodating significant disorder, implying diverse local environments around various cations.
An analysis of atomic bond distribution through AIMD simulations reveals distinct features in A-O pairs and a preference for SRO in A-A pairs at ambient temperature.  It is expected that all A-O couples will become blurred, while Mg-Mg pairs will exhibit a higher degree of diffusion as they approach a melting state at 3000 K.
Local atomic environments are further investigated by `large-box' RMC simulations, fitting neutron total scattering and EXAFS data simultaneously. Based on `small-box' fitting, AIMD, and RMC analysis results, the observed J14 structure suggests the existence of individualized local environments for cations, minor trends of preferred SRO for identical cations, and correlation length of LD.
These comprehensive findings offer invaluable insights for advancing the development of novel entropy-stabilized oxides and related compositionally complex phases with tailored properties for diverse materials applications.

\begin{acknowledgments}
This work is primarily supported by the National Science Foundation CAREER Award under NSF-DMR-2145174, from the Solid State and Materials Chemistry Program in the Division of Materials. This research used the NOMAD beamline at the Spallation Neutron Source, a DOE Office of Science User Facility operated by the Oak Ridge National Laboratory.  The computing resources were made available through the VirtuES project as well as the Compute and Data Environment for Science (CADES) at the Oak Ridge National Laboratory supported by the Office of Science of the U.S. Department of Energy under Contract No. DE-AC05-00OR22725, and Uninett Sigma2 supported by the University of Oslo (project no. NN4604K). 
This research used the beamline 12-BM of the Advanced Photon Source, a U.S. Department of Energy (DOE) Office of Science user facility operated for the DOE Office of Science by Argonne National Laboratory under Contract No. DE-AC02-06CH11357. C.M.R and J-P. M. acknowledges support from ARO under contract W911NF-14-0285 and NSF Materials Research Science and Engineering Centers (MRSEC) under Award No. DMR 2011839.
\end{acknowledgments}

\section{Materials and Methods}
\subsection{Materials and X-ray absorption spectroscopy}
Single phase J14 powder was synthesized using conventional solid-state methods as described in the following references.\cite{rost2015entropy,rost2017local} Extended X-ray absorption fine structure (EXAFS) spectra were collected at beamline 12-BM-B at the Advanced Photon Source, Argonne National Laboratory (Argonne, IL). Samples were prepared for transmission mode measurements on the Co, Ni, Cu, and Zn K-edges (7709 eV – 9659 eV), respectively. Data was processed and analyzed using the Demeter3 suite of programs for XAFS analysis.\cite{ravel2005athena} Individual scattering paths for each absorber were generated through custom .cif file inputs using FEFF\cite{rehr2009ab} and fit to the magnitude of the Fourier transformed EXAFS ($\chi$(k)) signal, shown in supplementary Figure S1. Uncertainties were generated through the least squares output of each fit, summarized in Table S1. The specific details of this data collection and analysis are published elsewhere.\cite{rost2017local,rost2016entropy} Using this data, EXAFS input files including initial cluster model paths, amplitude reduction factors, and edge energy shifts, each relating to absorber type were created for further use with RMCProfile simulations. 

\subsection{Neutron total scattering}
Neutron total scattering measurement of J14 was conducted at ambient temperature on the NOMAD diffractometer,\cite{neuefeind2012nanoscale} at the Spallation Neutron Source (SNS). Total scattering data was reduced using the ADDIE software suite\cite{mcdonnell2017addie}. Rietveld analysis was applied to neutron diffraction data using TOPAS v6 software \cite{coehlo2016topas} with the rocksalt cubic structure for J14. The refinement process included adjusting scale factor, profile parameters, unit cell parameters, background, atomic positions, and isotropic atomic displacements. 
Following the obtaining the experimental PDF data, further steps involved least-squares `small-box' modeling using PDFgui\cite{farrow2007pdffit2} and performing `large-box' RMC refinements using RMCProfile software\cite{tucker2007rmcprofile}. A 15$ \times $15$ \times $15 supercell from rocksalt cubic ($Fm\overline{3}m$) structure containing 27000 atoms was used in the RMC refinements by fitting PDF D(r)\cite{keen2001comparison,tucker2007rmcprofile}, the scattering function F(Q), neutron diffraction Bragg patterns and EXAFS simultaneously under a bond-valence-sum (BVS) constraint. Furthermore, a random distribution of five A-sites cations without atom swap was assumed and compared with the case with the atom swapping algorithm. The simulations ran on one 3.5GHz core CPU, generating more than 1.5$ \times 10^7$ moves for each RMC refinement run. \\

\subsection{Computational methods}
The special quasirandom structures (SQS)\cite{zunger1990special} for mimicking random solid solutions of J14 were prepared by mcsqs utility in the Alloy Theoretic Automated Toolkit (ATAT) \cite{van2009multicomponent,van2013efficient}. The Metropolis Monte Carlo (MMC) algorithm was utilized to perform random exchanges of atomic positions of A-site atoms, to create a random distribution of M-site disorder in periodic superlattices\cite{van2013efficient}. 
Additional density functional theory (DFT) calculations and $Ab$ $initio$ molecular dynamics (AIMD) were performed based on the \textit{Vienna Ab initio Simulation Package} (VASP) code \cite{kresse1996efficient,kresse1999ultrasoft} using the PBEsol functional \cite{perdew2008restoring,perdew1996generalized} which improves the prediction of lattice constants. 
The standard PBE PAW potentials Mg(2p$^6$3s$^2$), Co(3d$^7$4s$^2$), Ni(3p$^6$3d$^9$4s$^1$),  Cu(3p$^6$3d$^{10}$4s$^1$), Zn(3d$^{10}$4s$^2$) and O(1s$^2$2p$^4$) were used with the plane waves expanded up to cutoff energy of 560 eV. Additionally,  selected Hubbard U term values of 5.0, 5.1, 4.5, and 7.5 were applied to Co, Ni, Cu, and Zn to better describe the localized 3d electrons, as reported in \cite{rak2016charge}.
Brillouin zone integrations were completed with k-points density of at least 20 k-points Å$^{-1}$ for special SQS supercells in this work.  
A plane wave cutoff energy of 560 eV was used and all the structures were relaxed until the atomic forces were below 0.001 eV$/$\AA. More structures and calculation details will be discussed with the results. AIMD simulations using the NVT ensemble \cite{windiks2003massive} were carried out via the algorithm of Nos\'e \cite{nose1984unified} to control the temperature oscillations during the DFT calculations. Brillouin zone integration was completed on a single gamma-centered k-point and the cut-off energy was reduced to 450 eV for all AIMD calculations. Each AIMD simulation was run for more than 5000 steps at 300K with a time-step of 2 fs. The last 3000 steps were used to generate the time-averaged structure for each configuration.

\clearpage

\section{Supplementary Materials}
Supplementary material for this article is available online


\nocite{*}
\bibliography{ESO}

\end{document}